**Enhancing thermal stability of solution-processed small molecule semiconductor thin films using a flexible linker approach.**

*Andrea Gasperini, Xavier Jeanbourquin, Aiman Rahmanudin, Xiaoyun Yu and Kevin Sivula\**

A. Gasperini, X. Jeanbourquin, A. Rahmanudin, X. Yu, Prof. K. Sivula
Laboratory for Molecular Engineering of Optoelectronic Nanomaterials, École Polytechnique Fédérale de Lausanne (EPFL), Station 6, 1015 Lausanne, Switzerland.
E-mail: kevin.sivula@epfl.ch

Solution-processable small molecule organic semiconductors have recently emerged as promising materials for application in low-cost thin-film transistors,[1] light emitting diodes[2] and photovoltaic cells.[3] In contrast to the more-common conjugated polymer semiconductors, molecular semiconductors offer the advantages of straightforward purification, structural precision, and batch-to-batch consistency.[4] However, the strong tendency of conjugated molecular organic semiconductors to self-assemble into crystalline domains results in drawbacks like dewetting, poor thin film formation due to low solution viscosity, unpredictable crystalline domain sizes, and grain boundaries and that confound the morphological control and charge transport in devices fabricated from these materials. In particular, the need to control the thin film morphology is especially important for application in bulk heterojunction (BHJ) photovoltaics where the domain size between the electron donating and electron accepting phases must be precisely controlled to afford high power conversion efficiency.[5] For BHJ devices prepared with molecular semiconductor donors and the common small-molecule fullerene derivative, phenyl-$C_{61}$-butyric acid methyl ester ($PC_{61}BM$), as the acceptor the crystallization of the donor phase has been recently shown to be the driving force for blend separation.[6] Thus developing strategies to afford control over donor crystallization is an important goal. While volatile additives like diiodooctane[7-9] and more recently non-volatile nucleation promotors[10, 11] or insulating polymers[12] have been employed to offer control over the over the thin-film formation and crystalline domain size, these approaches do not nominally act on the donor phase nor do they address the intrinsic stability of the thin film morphology. Indeed, enhancing the long-term morphological stability of BHJ photovoltaics is one of the major remaining challenges in the field.[13, 14] Specifically, the inherent thermodynamic immiscibility of the donor and acceptor phases drives phase separation and eventually decreases device performance, even at normal operating temperatures.[15]

Given the drawbacks of molecular organic semiconductors, a "flexible linker" approach,[16] wherein well-defined conjugated segments are covalently linked with flexible aliphatic chains into polymeric materials, could reasonably provide a route to offer control over molecular donor crystallization and improve morphological stability in thin film devices, while also maintaining the consistency of the conjugated moiety. Herein, we demonstrate this flexible-linker strategy for molecular organic semiconductors for the first time using a common small molecule donor, 3,6-Bis(5-(benzofuran-2-yl)thiophen-2-yl)-2,5-bis(2-ethylhexyl)pyrrolo[3,4-c]pyrrole-1,4(2H,5H)-dione, coded as DPP(TBFu)$_2$.[17] When the flexibly-linked material is employed as an additive in thin films, a unique level of control is observed over the morphology and the thermal stability of pure donor films in transistor devices and in blends with PC$_{61}$BM for BHJ photovoltaics.

The synthesis of the flexibly-linked DPP(TBFu)$_2$ derivative [coded FL-DPP(TBFu)$_2$] is shown schematically in **Figure 1a**. The Stille-coupling polycondensation between the brominated diketopyrrolopyrrole core and the stannylated benzofuran dimer followed by standard purification and finally preparatory size exclusion chromatography afforded the pure polymer with a number average molecular weight, M$_n$, of 10 kDa and a PDI of 1.4 (n ≈ 11-15) in the hundred-milligram scale. Full synthetic procedures and detailed characterization are given in the Experimental Section and Supporting Information. To gain a first insight into the crystallization behavior of the FL-DPP(TBFu)$_2$ polymer we performed differential scanning calorimetry (DSC). The neat material did not exhibit a detectable glass transition nor enthalpic transitions associated with melting/crystallization in the temperature range from 50-240°C. However, when combined with the parent DPP(TBFu)$_2$ small molecule, a significant change in the crystallization occurred. Cooling curves obtained after melting blends of the FL-additive (at a weight fraction, $f_{FL}$) in the small molecule are shown in Figure 1b. A strong dependency on the DPP(TBFu)$_2$ crystallization temperature, $T_c$, upon addition of the FL-additive is clearly observed. With $f_{FL}$ at only 1 wt% a pronounced shift of the exothermic crystallization to higher temperatures is noted compared to the neat small molecule ($f_{FL} = 0$). This observation is consistent with the behavior of a nucleation promotor,[10, 11, 14] suggesting the FL-additive acts to reduce the driving force needed to nucleate stable nuclei of DPP(TBFu)$_2$. The maximum shift of the crystallization onset was found to be $\Delta T_c = 10$ °C with $f_{FL} = 5$ wt%, which is consistent in magnitude with recent reports of small molecule, non-conjugated nucleation promoters.[14] For higher $f_{FL}$ the onset of crystallization returns to lower temperatures and the enthalpy of the crystallization decreases suggesting the complete disruption of the ability of the blend to crystallize. We note that for $f_{FL} = 1$ and 5 wt%, comparable enthalpies of crystallization are observed ($\Delta H_c = 34.7$ and 30.1 J g$^{-1}$, respectively) with respect to the neat material (39.9 J g$^{-1}$) suggesting that the overall degree of crystallinity is not strongly reduced despite the addition of the amorphous FL-DPP(TBFu)$_2$. In addition we note that despite the drastic change in the crystallization behavior, only a small change was observed on the melting behavior (see Supporting information Figure S1) consistent with other reports of nucleation promoters.[18]

The effect of our FL strategy on the on the charge transport properties of solution-cast thin films was next investigated in bottom-contact/bottom-gate field-effect transistors. Devices prepared with the neat FL material showed no field-effect mobility ($\mu_{FET}$) confirming that the absence of crystalline ordering (as indicated by the DSC results) prevents long-range charge carrier transport. However, as-cast films of the FL material blended with DPP(TBFu)$_2$ gave measureable $\mu_{FET}$ for a wide range of $f_{FL}$, and for $f_{FL}$ up to 5 wt% the values were comparable with the neat DPP(TBFu)$_2$ film with $\mu_{FET} \approx 1\text{-}2 \times 10^{-5}$ cm$^2$ V$^{-1}$s$^{-1}$ (see Figure S2, Supporting Information). We note that the values of the as-cast $\mu_{FET}$ are similar to previous reports of the as-synthesized small molecule (which contains a mixture of stereoisomers).[19] As annealing is typically performed to increase the crystalline ordering and the charge carrier transport,[20] we next investigated the effect of the annealing temperature on transistors prepared with neat DPP(TBFu)$_2$ ($f_{FL} = 0$) and for $f_{FL} = 1$ and 5 wt%. The results are shown in **Figure 2a**. Annealing at 50°C for 30 min improved $\mu_{FET}$ in all cases, however, subsequent annealing at higher temperatures causes a decrease in $\mu_{FET}$ for neat DPP(TBFu)$_2$ devices. In contrast, the observed temperature dependence is quite different in the presence of the FL additive. The measured $\mu_{FET}$ increases slightly as a function of temperature until 150°C for the 1 wt% devices while the 5 wt% devices remain fairly constant from 70°C - 150°C, suggesting a more robust charge carrier transport network is created when adding the FL-DPP(TBFu)$_2$.

To better establish the thermal stability of the active layer, a set of devices, prepared by annealing at 100°C to have similar initial performance, were further subjected to a long-term thermal stress, periodically quenching the films to 30°C for testing. The measured $\mu_{FET}$ is reported against the annealing time at 100°C in Figure 2b. While $\mu_{FET}$ drops an order of magnitude for $f_{FL} = 0$, a considerably smaller decrease is observed when $f_{FL} = 1$ wt%, and notably at 5 wt% the performance remains constant. Visualizing the active layer morphology with atomic force microscopy (AFM) gives further insight into the origin of the improved thermal performance of the transistors with the added FL-DPP(TBFu)$_2$. As-cast films with 0-5 wt% of the FL additive all exhibit a similar fibril-type morphology (Figure S3, SI). After the extended thermal stress test at 100°C for a total of 3 hours, a drastic difference is observed between the topology of the films as shown in Figure 2c-e. The neat (0 wt%) device exhibits small circular domains (ca. 100-200 nm) and a few longer shards. The disappearance here of the initial fibril type morphology may be due to a dewetting from the substrate or reasonably caused by the formation of defects during the rapid quenching process at the semiconductor dielectric interface[21] which propagate and alter the domain morphology. Indeed, the thermal annealing of solution-processed molecular semiconductor films has recently been shown to induce the formation of trapping states, likely at grain boundaries, that reduce charge carrier transport.[22] In stark contrast to the neat DPP(TBFu)$_2$ devices, when $f_{FL}$ = 1 or 5 wt% films exhibit large banded features 200-500 nm in width and more than microns in length after the extended annealing test. In addition, the roughness of the film decreases considerably with $f_{FL}$ = 5 wt% (see height profiles Figure S4, SI).

The transistor device thermal stability data together with the evolution of the morphology demonstrate that the FL-DPP(TBFu)$_2$ actively participates in the stabilization of the thin-film charge transport network. In contrast to the typical behavior of a nucleation promotor, which increases the number of nucleation events thus leading to a smaller crystalline domain size,[11] the FL additive also appears that this point to promote larger crystalline domains. Given the polymeric structure of our FL additive one possible explanation for this behavior would be that it is acting as a tie molecule to bridge adjacent crystal domains, effectively locking-in the active layer morphology. Given the relatively short length of the FL polymer chain (ca. 40 nm) compared to the observed domain size, and the amount of FL additive employed, this explanation would be possible if the FL-additive is positioned at the crystal grain boundaries and the polymer chain bridged two adjacent domains. This is reasonable as the FL-additive is likely positioned at the grain boundary due to its action as a nucleation promotor. However, we do not suspect that the FL-additive can act as a traditional tie-molecule, which typically would link many crystallite domains. Another aspect that may affect the formation of a more stable charge transport network is the nucleation of a different crystal form of the DPP(TBFu)$_2$ via the FL additive. Indeed polymorph formation has been observed with stereoisomers of DPP(TBFu)$_2$ and are known to have an effect on the charge transport.[18] Analysis of thick films (deposited on SiO$_2$ and annealed at a high temperature, 190 °C, below the melting point to maximize crystallinity) by grazing incidence wide angle x-ray scattering (GIWAXS, See Figure S5, supporting information) supports the notion that a change in crystal formation could play a role. While at $f_{FL}$ = 0 the diffraction peaks match previously reported films annealed at low temperature (80°C),[10] distinctive Bragg rods are observed when $f_{FL}$ = 1 wt% suggesting a thin film polymorphism.[23] Analysis of the films annealed at 190°C by AFM supports this by indicating a clear change from fiber-like features (500 nm in width) observed in the control film (Figure S5a) to wide platelets observed with $f_{FL}$ = 1 wt% (Figure S5c). While the complete understanding of the complex role of the FL- DPP(TBFu)$_2$ on the thin film crystallinity is undoubtedly beyond the scope of this initial demonstration of the flexible linker concept, the strong effect of the FL additive on the morphology and performance of donor-only films is clear.

Given the common use of DPP(TBFu)$_2$ as an electron-donor small-molecule blended in bulk heterojunction (BHJ) photovoltaic cells with fullerene-based electron acceptors, we next probed the effect of the inclusion of the FL additive in DPP(TBFu)$_2$:PC$_{61}$BM solar cells. Standard ITO/PEDOT:PSS/BHJ/Al devices were fabricated by typical procedures as described in the experimental section. We compared performance of devices with $f_{FL}$ = 0 - 5 wt% (based on the total mass of the active layer in DPP(TBFu)$_2$:PC$_{61}$BM blends at a ratio of 3:2) and found, for devices tested without annealing, the photovoltaic figures of merit, J$_{sc}$, V$_{oc}$ and FF were not affected by the presence of the FL additive (see Supporting Information, Figure S6) and were consistent with previous reports with "as-cast" active layers.[17] The invariance of V$_{oc}$ supports the notion that the presence of the additive does not change the HOMO level of the acceptor phase, which is reasonable considering the nature of the conjugated segment is unchanged. However, the unchanging J$_{sc}$ and FF values of the as-cast devices with $f_{FL}$ = 0 - 5 wt% suggest that the FL-additive does not have a significant effect on the BHJ morphology during the initial film formation (spin casting from chlorobenzene) despite its established activity as a nucleation promotor for the donor phase.

Annealing devices with $f_{FL} = 0$ at 100°C for 10 min gave an average device power conversion efficiency, PCE, of 3.4 ± 0.3%. We note that this value is slightly lower than state-of-the-art values obtained when using PC$_{71}$BM due to the expected lower J$_{sc}$. Devices with $f_{FL} = 0.5$ and 1.5 wt% exhibited equivalent performance when the annealing time was optimized. Interestingly, we found it necessary to extend the annealing to 1 hour to afford PCE equal to the $f_{FL} = 0$ case. Optimized "best-case" J-V curves for $f_{FL} = 0 - 5$ wt% are shown in **Figure 3a** and the average PCE as a function of the annealing time at 100°C is shown for $f_{FL} = 0$ and 0.5 wt% in Figure 3b. When $f_{FL} = 3$ or 5 wt% the best-case J-V performance is significantly decreased, exhibiting lower FF and J$_{sc}$, which indicates an increased recombination and poorer charge carrier transport. Remarkably, the thermal stability (Figure 3b) of the device performance was found to be significantly different between the $f_{FL} = 0$ and 0.5 wt%. After only 3 hours of thermal stress at 100°C a decrease in the PCE from 3.4% to 2.3% is observed when $f_{FL} = 0$. This observed trend, common for many systems employing PC$_{61}$BM, is primarily attributed to a decrease in FF and J$_{sc}$ due to the phase segregation of the PC$_{61}$BM from the donor phase. In contrast, device performance remains stable at 3.2% when $f_{FL} = 0.5$ wt% even after 22 hours of annealing at 100°C. This suggests a different evolution of the microstructure of the BHJ during the thermal stress test.

Insight into the origin of the improved thermal stability of the donor:acceptor blends with the FL-additive was next sought by examining the morphology of the active layer with AFM and nanomechanical mapping[24, 25] to identify donor-rich and acceptor-rich domains. Indeed, as the Young's modulus (YM) of neat PC$_{61}$BM (ca. 17 GPa) and neat DPP(TBFu)$_2$ (ca. 3 GPa) are significantly different (see Supporting Information Figure S7), YM mapping can give information on the domain composition. Combined topographical and YM map results are shown in **Figure 4**. The active layer topography after annealing at 100°C for 10 min (unconfined by the Al top electrode) is shown in Figure 4a (left panel), and exhibits rod-shaped domains 100-200 nm in size when $f_{FL} = 0$ (similar to previous reports[17]). We note that the topography of as-cast BHJ films when $f_{FL} = 0 - 5$ wt% are similar, except with slightly larger rod-shaped domains for $f_{FL} = 0.5$ and 1.5 wt% (see Figure S8 supporting information). The YM map corresponding to the topography image of Figure 4a is displayed on the right of the figure. Here domains of low YM, corresponding roughly to the rod-shaped domains, are distinguishable from areas of higher modulus, confirming that these rod domains are donor-rich and further suggesting that the donor and acceptor are mixed at a length scale of ca. 100 nm. For films with no FL-additive ($f_{FL} = 0$) that have been annealed for 22 hours at 100°C, the topology (Figure 4b) shows the absence of rod-shaped domains and the presence of circular domains similar to the donor only film shown in Figure 2c. The corresponding YM map gives values < 6 GPa suggesting that the surface of the film is composed primarily of DPP(TBFu)$_2$ (i.e. no PC$_{61}$BM-rich domains are present). Indeed, micron-sized PC$_{61}$BM-rich domains are observed by AFM, YM mapping, and by optical microscopy (See Figure S9) suggesting that the PC$_{61}$BM has segregated from the DPP(TBFu)$_2$ on a length scale too-large for BHJ operation. We note that the performance of the corresponding photovoltaic device ($f_{FL} = 0$) after annealing for 22 hours likely stabilizes at a non-zero value due to the effect of the morphological confinement brought by the Al top contact.[26] With the FL-additive present in the BHJ, a considerably different morphology is observed after thermal stress for 22 hours (Figure 4c). In this case the topography shows a comparatively smother film with long rod-shaped domains similar to those observed in the neat films (Figure 2d-e). The YM map indicates that these rod-shaped regions exhibit a lower modulus consistent with the DPP(TBFu)$_2$. Moreover, we note that Young's moduli higher than 10 GPa are also measured in high proximity to the rod-shaped domains, indicating that PC$_{61}$BM-rich domains remain present in the film with the FL-additive present after thermal treatment for 22 hours. This considerable difference in behavior of the morphology is indeed consistent with the observed difference in device performance and suggests that the FL-additive strategy is effective at extending the morphological stability of BHJ molecular solar cells.

In addition, we note that presence of the comparatively longer rod-shaped crystalline domains in observed in the morphological study suggest an apparent increase in donor domain size when the FL-additive is present in blends with $PC_{61}BM$ and optimized thermal annealing is performed. However, we note that out-of-plane XRD measurements indicate that the donor has similar crystallinity in the blend film without the FL-additive after 10 min annealing compared to the $f_{FL}$ = 0.5 wt% blend film when annealing for 1 hour (See supporting information Figure S10) therefore no increased overall crystallinity of the donor phase is observed using the FL-additive consistent with the DSC results which suggest a slight decrease in donor crystallinity. However, supposing that the longer rod-shaped donor domains observed in stabilized BHJ blends with the FL-additive could afford improved charge carrier transport over longer distance, we hypothesized that a greater difference in device performance (with and without the FL-additive) would be observed if thicker active layers were employed. To provide preliminary evidence to support this notion, we increased the thickness of the active layer by 50% (i.e. from 90 nm to 140 nm). After optimized annealing at 100°C, devices with $f_{FL}$ = 0 gave a 40% lower PCE compared to devices with $f_{FL}$ = 0.5 wt%. As expected, the difference was due to a drop in FF and $J_{sc}$ which are found to be, respectively, 0.34 and 7.5 mA cm$^{-2}$ (for $f_{FL}$ = 0), compared to 0.45 and 9.9 mA cm$^{-2}$ (for $f_{FL}$ = 0.5 wt%, see Supporting Information Figure S11). Even though significant PCE improvement was not observed over the thinner cells, this result suggests that in thicker BHJ films—where increased crystallinity is required to afford efficient charge extraction—the FL-additive can support a more intimate mixing of the donor and acceptor, while still providing suitable crystallinity for charge transport. Indeed we note that for BHJ films annealed for one hour with or without the FL additive, films without the additive exhibited increased donor crystallinity (consistent with the donor-driven crystallization in this system, See Figure S10

Supporting Information) which drives the phase segregation and the lower $J_{SC}$. The improved BHJ thermal stability and lower comparative donor crystallinity can be rationalized again by the ability the FL-additive to promote the nucleation of the donor phase. The presence of the amorphous FL-additive at the grain boundaries of the donor crystallites could reasonably maintain a stable donor transport network while also providing amorphous regions for PCBM, thereby limiting its exclusion. This combination of beneficial characteristics is unique among additives to bulk heterojunctions. We cannot, however, exclude the possibility of donor polymorph formation playing a role in the BHJ stability. Indeed, while out-of-plane XRD measurements do not indicate any considerable difference in the observed diffraction peak, a difference is not expected even with polymorphs as reported DPP(TBFu)$_2$ polymorphs exhibit similar scattering vectors for the main peak.[19] Further analysis of the BHJ crystallinity is a point of interest given these initial results.

Importantly, we note that the FL-additive strategy presented here is significantly different than recently-reported approaches to reactively cross-link the donor and acceptor after BHJ formation.[27, 28] While these crosslinking approaches have been shown to improve the thermal stability of BHJs when polymer donor phases are used, as a drawback they rely on the uncontrolled reaction of functional groups that likely add electronic defects to the film. In addition, we emphasize that the observed behavior of the FL-additive is also distinct from simply blending in insulating (inert) polymer additives, in molecular bulk heterojunctions.[12] While this approach offers some control over the thin film formation, since the inert polymer segregates from the BHJ, it likely does not improve thermal stability.

Overall the device results and morphological investigations presented here demonstrate that the strategy of linking molecular semiconductors with flexible aliphatic chains into polymers that possess defined conjugated segments, but extended covalent connectivity is a viable way to control the crystalline behavior of the solution-processed molecular semiconductor thin films. We have demonstrated that our new class of polymeric additive can promote donor nucleation in neat films and BHJ blends to improve thermal stability and long range charge transport. We have also shown, for the first time, that small molecule BHJ solar cell thermal stability can be drastically improved by including our flexibly-linked additive. Since the flexible linker approach is straightforward and can be applied to a wide variety of different conjugated small molecules, we believe that this strategy can have broad application in controlling and stabilizing the active layer morphology in molecular BHJ PVs, OLEDs, and TFTs.

**Experimental Section**

*Material synthesis and basic characterization:* The DPP(TBFu)$_2$ and PC$_{61}$BM small molecules were prepared using standard procedures.[17] The FL- DPP(TBFu)$_2$ polymer was prepared using a procedure adapted from previous work.[16] See supporting information for full synthetic details as well as basic material characterization.

*Transistor fabrication and testing*: Bottom contact field-effect transistors (FETs) were fabricated on pre-patterned test substrates (Fraunhofer Institute for Photonic Microsystems) whose source and drain contacts were composed of a 30 nm thick gold layer on top of a 10 nm thick titanium adhesion layer. A 230 nm thick silicon oxide was used as gate dielectric and n-doped silicon wafer as the substrates and gate electrode. The channel length (L) was 10 or 20 μm and channel width (W) was 10 mm. The transistor substrates were cleaned by sonication in acetone and isopropanol at RT for 15 min in each solvent. After drying under nitrogen, the substrates were subsequently exposed to a nitrogen plasma for 30 min. Films of ~60 nm thickness were spin-coated from 10 mg mL$^{-1}$ DPP(TBFu)$_2$ + FL-DP(TBFU)$_2$ solutions in CHCl$_3$ at 500 rpm (for 1 min). The solutions were prepared by dissolution at 55 °C for 1h under continuous stirring. All solutions and films were prepared in an argon atmosphere. Unannealed films were left overnight under vacuum (0.01 mbar) to remove solvent before testing. Electronic testing of the FETs was carried out in a nitrogen atmosphere using a custom-built probe station and a Keithley 2612A dual-channel source measure unit. Samples were annealed in a nitrogen atmosphere and quenched to room temperature before electronic testing. The field effect mobility was extracted from the saturation region as previously described.[16] Six devices were measured at each condition and the average values are reported.

*Thermal characterization:* Melting temperatures, $T_m$, and enthalpies of fusion, $\Delta H_f$, were measured with a Perkin Elmer DSC8000 differential scanning calorimeter calibrated with indium and zinc, using a scanning rate of 10°C min$^{-1}$. Samples were prepared by drop casting and slow evaporation of the solvent at 40°C under argon atmosphere before the measurement. $\Delta H_f$ was calculated by integration of the melting endotherm.

*Solar cell fabrication and testing*: Solar cells were fabricated with a 35-nm layer of PEDOT:PSS (Ossilla M 121 Al 4083) deposited and annealed (100°C) on a glass substrate patterned with 300 nm of ITO. The BHJ active layer was spin-cast at 3000 rpm from a solution of DPP(TBFu)$_2$ and PC$_{61}$BM in chloroform at a total solids concentration of 20-30 mg mL$^{-1}$. The active layers were determined to be approximately 90-nm thick using a Bruker Dektak XT profilometer. An 80-nm-thick aluminum cathode was deposited (area 16 mm$^2$) by thermal evaporation (Kurt J. Lesker Mini-SPECTROS). Electronic characterization was performed under simulated AM1.5G irradiation from a 300W Xe arc lamp set to 100 mW cm$^2$ with a calibrated Si photodiode (ThorLabs). Current-voltage curves were obtained with a Keithley 2400 source measure unit. Device fabrication was performed under an argon atmosphere and testing was performed under in a nitrogen filled glovebox.

*Force peak mapping:* Peak force measurements were recorded in ambient conditions using a Cypher S AFM from Asylum Research. A silicon tip (AC160TS) with a theoretical spring constant of 42 N m$^{-1}$ was used. The experimental spring constant was obtained using the GetReal$^{TM}$ automated probe calibration feature of Asylum Research software. Over all the scans, a peak force of 100 nN was kept constant for all the samples. The reduced Young's modulus was obtained by fitting the retraction curve with the Derjaguin, Muller, Toropov (DMT) model, which accounts for the adhesion between the tip and the sample surface.

*Diffraction characterization*: GIWAXS measurements were carried out at ANKA synchrotron in KIT on beam line PDIFF. The incident energy was 11 keV. The films were illuminated at a constant incidence angle of 0.15° using a 2D-area detector. The distance between the sample and the 2D camera was set at 180 mm. Out of plane, specular X-ray scattering was performed with a standard Cu Kα laboratory source (0.15418 nm) on a Bruker D8 Advanced goniometer using grazing incidence conditions with an incident angle of 0.3°, a scan speed of 0.05 deg min$^{-1}$, and a Lynx Eye linear detector.

**Supporting Information**
Supporting Information is available from the Wiley Online Library (http://dx.doi.org/10.1002/adma.201501826) or from the author.

**Acknowledgements**
We thank the Swiss National Science Foundation (Project 200021_144311) and the European Research Commission (project 336506 – CEMOS) for financial support. Dr. Stephen Doyle, at the ANKA Beamline is gratefully acknowledged for assistance with the 2D GIWAXS measuements.

**References**

[1]     S. Allard, M. Forster, B. Souharce, H. Thiem, U. Scherf, *Angew. Chem. Int. Ed.* **2008**, *47*, 4070.

[2]     K. S. Yook, J. Y. Lee, *Adv. Mater.* **2014**, *26*, 4218.

[3]     Y. Lin, Y. Li, X. Zhan, *Chem. Soc. Rev.* **2012**, *41*, 4245.

[4]     B. Walker, A. Tamayo, D. T. Duong, X. D. Dang, C. Kim, J. Granstrom, T. Q. Nguyen, *Adv. Energy Mater.* **2011**, *1*, 221.

[5]     M. T. Dang, L. Hirsch, G. Wantz, J. D. Wuest, *Chem. Rev.* **2013**, *113*, 3734.

[6]     A. Sharenko, M. Kuik, M. F. Toney, T. Q. Nguyen, *Adv. Funct. Mater.* **2014**, *24*, 3543.

[7]     A. Viterisi, F. Gispert-Guirado, J. W. Ryan, E. Palomares, *J. Mater. Chem.* **2012**, *22*, 15175.

[8]     Y. M. Sun, G. C. Welch, W. L. Leong, C. J. Takacs, G. C. Bazan, A. J. Heeger, *Nat. Mater.* **2012**, *11*, 44.

[9]     L. A. Perez, J. T. Rogers, M. A. Brady, Y. M. Sun, G. C. Welch, K. Schmidt, M. F. Toney, H. Jinnai, A. J. Heeger, M. L. Chabinyc, G. C. Bazan, E. J. Kramer, *Chem. Mater.* **2014**, *26*, 6531.

[10]    A. Sharenko, N. D. Treat, J. A. Love, M. F. Toney, N. Stingelin, T. Q. Nguyen, *J. Mater. Chem. A* **2014**, *2*, 15717.

[11]    N. D. Treat, J. A. Nekuda Malik, O. Reid, L. Yu, C. G. Shuttle, G. Rumbles, C. J. Hawker, M. L. Chabinyc, P. Smith, N. Stingelin, *Nat. Mater.* **2013**, *12*, 628.

[12]    Y. Huang, W. Wen, S. Mukherjee, H. Ade, E. J. Kramer, G. C. Bazan, *Adv. Mater.* **2014**, *26*, 4168.

[13]    B. C. Schroeder, Z. Li, M. A. Brady, G. C. Faria, R. S. Ashraf, C. J. Takacs, J. S. Cowart, D. T. Duong, K. H. Chiu, C. H. Tan, J. T. Cabral, A. Salleo, M. L. Chabinyc, J. R. Durrant, I. McCulloch, *Angew. Chem. Int. Ed.* **2014**, *53*, 12870.

[14]    C. Lindqvist, J. Bergqvist, C. C. Feng, S. Gustafsson, O. Backe, N. D. Treat, C. Bounioux, P. Henriksson, R. Kroon, E. G. Wang, A. Sanz-Velasco, P. M. Kristiansen, N. Stingelin, E. Olsson, O. Inganas, M. R. Andersson, C. Muller, *Adv. Energy Mater.* **2014**, *4*, 1301437.

[15]    I. Cardinaletti, J. Kesters, S. Bertho, B. Conings, F. Piersimoni, J. D'Haen, L. Lutsen, M. Nesladek, B. Van Mele, G. Van Assche, K. Vandewal, A. Salleo, D. Vanderzande, W. Maes, J. V. Manca, *J. Photon. Energy* **2014**, *4*, 040997.

[16]    A. Gasperini, S. Bivaud, K. Sivula, *Chem. Sci.* **2014**, *5*, 4922.

[17]    B. Walker, A. B. Tomayo, X. D. Dang, P. Zalar, J. H. Seo, A. Garcia, M. Tantiwiwat, T. Q. Nguyen, *Adv. Funct. Mater.* **2009**, *19*, 3063.

[18]    W. J. Liu, H. L. Yang, Z. Wang, L. S. Dong, J. J. Liu, *J. Appl. Polym. Sci.* **2002**, *86*, 2145.

[19]     J. Liu, Y. Zhang, H. Phan, A. Sharenko, P. Moonsin, B. Walker, V. Promarak, T.-Q. Nguyen, *Adv. Mater.* **2013**, *25*, 3645.

[20]     Z. Li, X. Zhang, Y. Zhang, C. F. Woellner, M. Kuik, J. H. Liu, T. Q. Nguyen, G. Lu, *J. Phys. Chem. C* **2013**, *117*, 6730.

[21]     E. Gann, X. K. Gao, C. A. Di, C. R. McNeill, *Adv. Funct. Mater.* **2014**, *24*, 7211.

[22]     J. D. A. Lin, O. V. Mikhnenko, T. S. van der Poll, G. C. Bazan, T.-Q. Nguyen, *Adv. Mater.* **2015**, n/a.

[23]     S. C. B. Mannsfeld, M. L. Tang, Z. Bao, *Adv. Mater.* **2011**, *23*, 127.

[24]     D. Wang, T. P. Russell, T. Nishi, K. Nakajima, *ACS Macro Lett.* **2013**, *2*, 757.

[25]     D. Wang, F. Liu, N. Yagihashi, M. Nakaya, S. Ferdous, X. Liang, A. Muramatsu, K. Nakajima, T. P. Russell, *Nano Lett.* **2014**, *14*, 5727.

[26]     K. Sivula, Z. T. Ball, N. Watanabe, J. M. J. Frechet, *Adv. Mater.* **2006**, *18*, 206.

[27]     L. Derue, O. Dautel, A. Tournebize, M. Drees, H. Pan, S. Berthumeyrie, B. Pavageau, E. Cloutet, S. Chambon, L. Hirsch, A. Rivaton, P. Hudhomme, A. Facchetti, G. Wantz, *Adv. Mater.* **2014**, *26*, 5831.

[28]     C.-P. Chen, C.-Y. Huang, S.-C. Chuang, *Adv. Funct. Mater.* **2015**, *25*, 207.

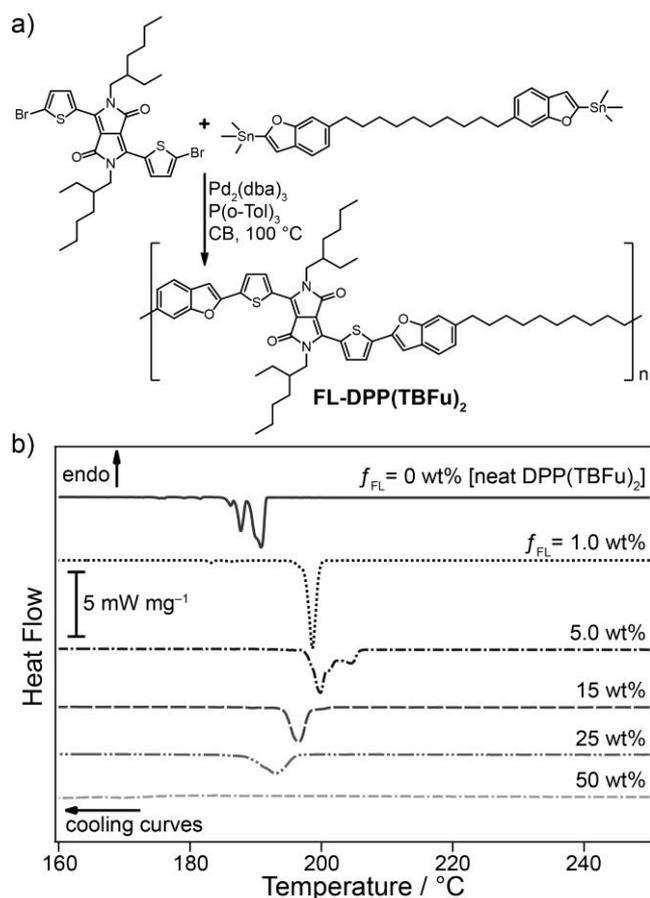

**Figure 1.** The synthetic scheme for FL-DPP(TBFu)$_2$ is shown in (a). Panel (b) shows the differential scanning calorimetry cooling curves for different amounts of FL-DPP(TBFu)$_2$ in DPP(TBFu)$_2$.

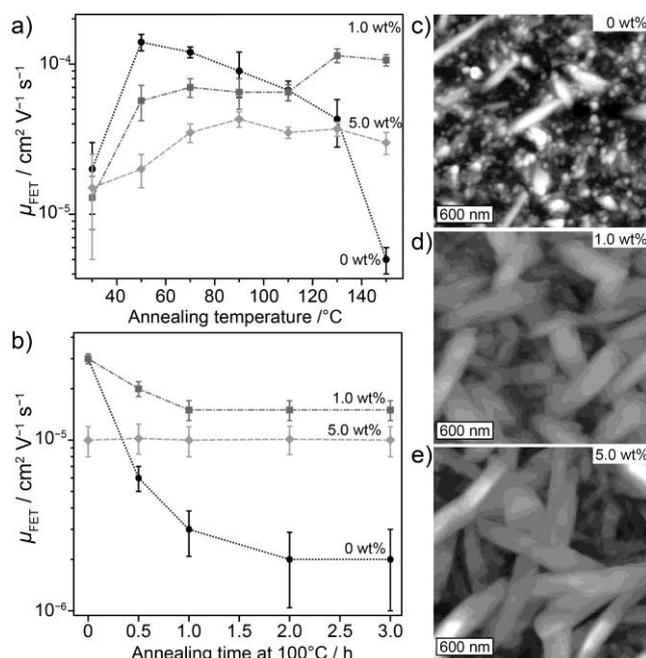

**Figure 2.** Thin film transistor performance and morphology. Panel (a) shows the average extracted field effect mobility as a function of thin film annealing temperature. Panel (b) displays the mobility as a function of annealing time at 100°C. Atomic force micrographs (c-e) show the topology of the thin film transistor active layer after 3.0 hours at 100°C.

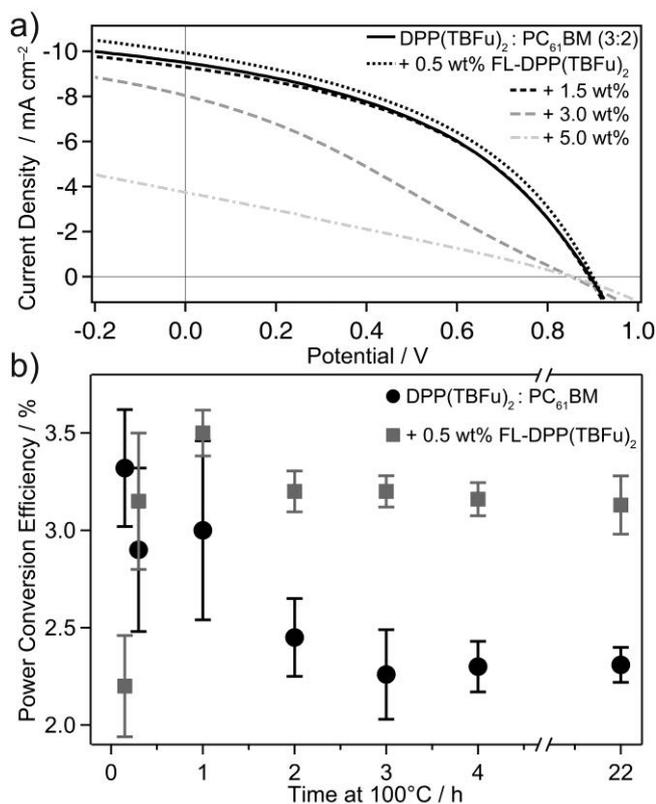

**Figure 3.** Bulk heterojunction device performance. (a) shows the best case J-V curves for devices prepared with a DPP(TBFu)$_2$:PC$_{61}$BM blend at a ratio of 3:2 and added FL-DPP(TBFu)$_2$. (b) shows the average power conversion efficiency as a function of the annealing time at 100°C.

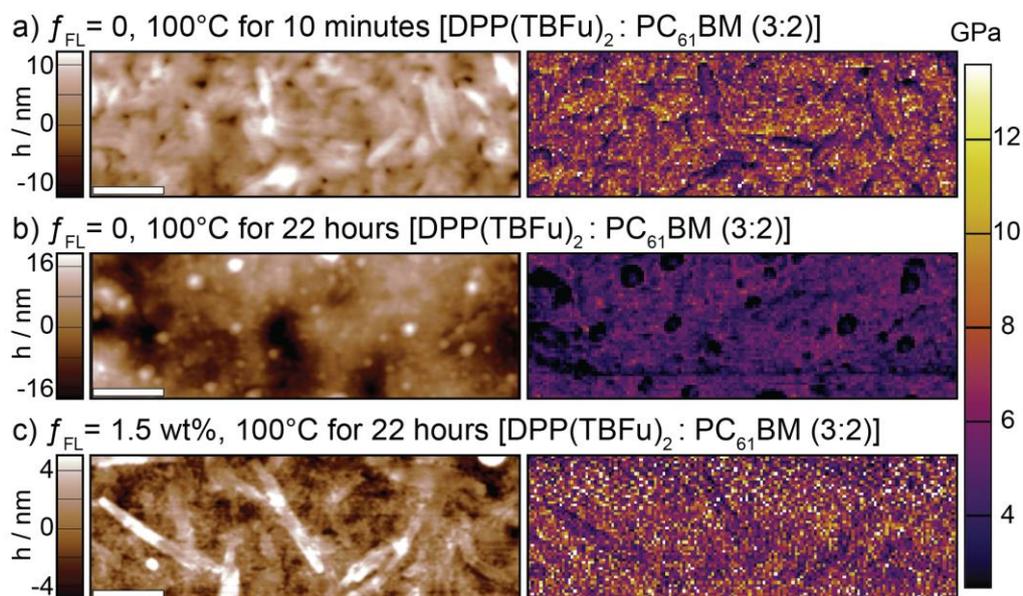

**Figure 4.** Bulk heterojunction topography (left panels) and Young's modulus mapping (right panels) of the corresponding area. The scale bars is 500 nm.

17